\documentclass{PoS}
\usepackage{amsmath}
\usepackage{amsfonts}
\usepackage{dsfont}

\newcommand{\expect}[1]{\left\langle #1 \right\rangle}
\newcommand{\id}{\mathds{1}}

\newcommand{\psibar}{\overline{\psi}}

\newcommand{\parR}{\mathcal{R}_5^{1,2}}

\newcommand{\distras}[1]{%
  \savebox{\mybox}{\hbox{\kern3pt$\scriptstyle#1$\kern3pt}}%
  \savebox{\mysim}{\hbox{$\sim$}}%
  \mathbin{\overset{#1}{\kern\z@\resizebox{\wd\mybox}{\ht\mysim}{$\sim$}}}%
}
\title{\vspace{-2cm}
{\small \normalfont \hfill DESY 13-206\\
  \hfill HU-EP-13/62\\
  \hfill SFB/CPP-13-88
\\}
\vspace{1cm} Computation of the chiral condensate using $N_f=2$ and $N_f=2+1+1$ dynamical flavors of twisted mass fermions\\}
\author{{K. Cichy $\,\,^{a,b}$, \speaker{E.
    Garcia-Ramos} $\,\,^{a,c}$, K.
Jansen$\,\,^a$, A. Shindler $\,\,^{d}$\\
\llap{$^a$} NIC, DESY, Platanenallee 6, D-15738 Zeuthen, Germany\\
\llap{$^b$} Adam Mickiewicz University, Faculty of Physics , Umultowska 85,
61-614 Pozna\'n, Poland\\
\llap{$^c$}  Humboldt Universit\"at zu Berlin,  Newtonstrasse 15 12489 Berlin,
Germany}\\
\llap{$^d$} IAS,IKP and JCHP, Forschungszentrum J\"ulich, 52428 J\"ulich, Germany\\
Email: \email{krzysztof.cichy@desy.de}, \email{elena.garcia.ramos@desy.de}, \email{karl.jansen@desy.de},   \email{a.shindler@fz-juelich.de}
}

\ShortTitle{$\Sigma$ for $N_f=2$ and $N_f=2+1+1$ dynamical flavors of twisted mass fermions.}
\abstract{We apply the spectral projector method, recently introduced
  by Giusti and L\"uscher, to compute the chiral condensate using
  $N_f=2$ and $N_f=2+1+1$ dynamical flavors of maximally twisted mass
  fermions. We present our results for several quark masses at three
  different lattice spacings which allows us to perform the chiral and
  continuum extrapolations. In addition we report our analysis on the
  $O(a)$ improvement of the chiral condensate for twisted mass fermions. 
We also study the effect of the dynamical strange and charm quarks by comparing our results for $N_f=2$
and $N_f=2+1+1$ dynamical flavors.
\begin{center}
\vspace*{1cm}
\includegraphics
[width=0.2\textwidth,angle=0]
{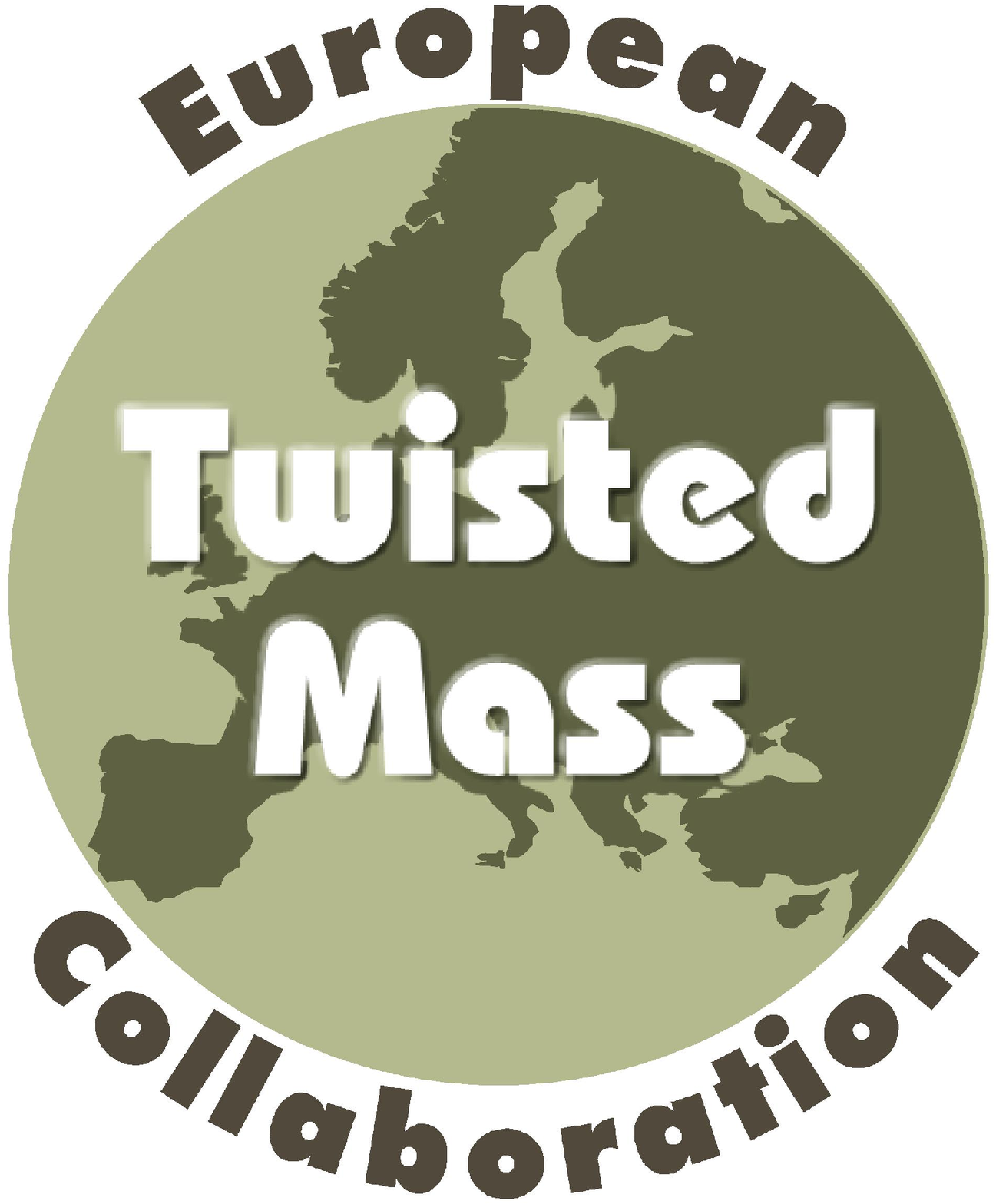}
\end{center}
}
\FullConference{31st International Symposium on Lattice Field Theory - LATTICE 2013\\
                July 29 - August 3, 2013\\
                Mainz, Germany}
\begin{document}

\section{Introduction}

The Banks-Casher relation \cite{Banks:1979yr} connects the low lying spectrum of the Dirac operator with the spontaneous chiral symmetry breaking in the following way
\begin{equation}
  \label{BCrelation}
  \lim_{\lambda\rightarrow0}\lim_{m\rightarrow0}\lim_{V\rightarrow\infty} \rho(\lambda,m)=\frac{\Sigma}{\pi}.
\end{equation}

Eq.~(\ref{BCrelation}) relates the chiral condensate $\Sigma$ to the spectral density $\rho(\lambda,m)$. The recently introduced method based on spectral projectors \cite{LuscherGiusti} offers a new strategy to compute spectral observables, such as the chiral condensate, in an affordable way. 
Moreover it allows us, via the connection to density chains, 
to compute this quantity using a representation which is free of short distance singularities and therefore leads to the correct continuum limit.

The integrated spectral density or mode number $\nu(M,m)$ is defined
as the number of eigenvalues $\lambda$ of the hermitian Dirac operator
$D^\dagger D$ below a certain threshold value $M^2$. To study the
renormalization and $O(a)$ cutoff effects properties of the mode
number it is advantageous to consider the spectral sums
$\sigma_k(\mu,m)$ which are directly related through the following
expression
\begin{equation}
  \label{rel} \sigma_k(\mu,m)=\int^{\infty}_0dM\;\nu(M,m)\frac{2kM}{(M^2+\mu^2)^{k+1}}.
\end{equation}

In particular, it is convenient to write the spectral sums $\sigma_k$
in terms of density chains like
\begin{equation}
  \label{specsum2}
   \sigma_3(\mu,m)=-a^{24}\sum_{x_1,\dots,x_5}\expect{P^+_{12}(x_1) P^-_{23}(x_2) P^+_{34}(x_3) P^-_{45}(x_4) P^+_{56}(x_5) P^-_{61}(0)},
\end{equation}
where $P_{ab}^\pm=\psibar_a\gamma_5\tau^\pm\psi_b$ are charged
pseudoscalar densities, $\tau^\pm$ are defined in flavor
space and $\mu$ is the twisted
mass. $\psibar_a=(\overline{u_a},\overline{d_a})$ represent isospin
doublets of 
twisted mass valence fermions. The index $a=1,\dots,k$
indicates the doublet; in this particular example we add 6 doublets to
the theory, which is the minimum number of flavors that still
guarantees the renormalizability as it was shown in \cite{LuscherGiusti}.

 Nevertheless, in the end, the mode number contains the same information as the spectral density and therefore it is directly linked to the chiral condensate as proposed in Ref.~\cite{LuscherGiusti}
\begin{equation}
  \label{SigmaGL} \Sigma_R=\frac{\pi}{2V}\sqrt{1-\left(\frac{\mu_R}{M_R}\right)}\frac{\partial}{\partial M_R}\nu_R(M_R,\mu_R).
\end{equation}
Notice that $\nu_R(M_R,m_R)=\nu(M,m)$, the mode number is a renormalization group invariant \cite{LuscherGiusti}.

It goes beyond the scope of this proceedings to explain in detail the
spectral projectors method and its implementation to compute the
chiral condensate among other observables. Thus, for further details
we refer to the original article Ref.\cite{LuscherGiusti} and also 
to Ref.~\cite{Cichy:2013gja} for the implemention of the setup used 
here. For a computation of the topological susceptibility in the same
set up using spectral projectors, following the method described in
Ref.\cite{Luscher:2010ik}, see Ref.~\cite{tsPoS}. In this 
contribution we present the continuum limit results obtained for the
chirally extrapolated condensate $\Sigma$ for $N_f=2$ and $N_f=2+1+1$
dynamical flavors of maximally twisted mass fermions
\cite{Frezzotti:2000nk, Frezzotti:2004wz, Shindler:2007vp,
  Frezzotti:2003xj} using the
spectral projectors method. In addition we discuss the $O(a)$
improvement of this quantity, since it is, in principle, possible that
contact terms that arise in the Symanzik expansion introduce $O(a)$
cut-off effects and spoil automatic $O(a)$ improvement. 

\section{$O(a)$ improvement of the chiral condensate}

The representation of the mode number and the spectral density of the 
Wilson operator through density chains correlator as in
Eq.~(\ref{specsum2}) allows to discuss the renormalization and improvement
properties of such quantities.
This is particularly important when computing the mode number using Wilson twisted mass
fermions at maximal twist. 
The maximal twist condition guarantees automatic $O(a)$ improvement of all physical
quantities~\cite{Frezzotti:2003ni}.
In fact one can show that the observable introduced in
Eq.~(\ref{specsum2}) is even under $\parR$ transformations given by
$\chi_i(x) \rightarrow i\gamma_5\tau^{1,2}\chi_i(x), \quad\overline{\chi_i}(x) \rightarrow \overline{\chi_i}(x)i\gamma_5\tau^{1,2}$,
where $\chi_i$ refers to the valence and sea twisted mass quarks.
Consequently the automatic $O(a)$ improvement obtained by tuning to maximal twist should apply. 

Density chains correlators are affected by short distance singularities and 
the integration over the whole space-time of such singularities generates additional $O(a)$ terms 
that could spoil the property of automatic $O(a)$ improvement.
In this section we argue that those terms vanish at
maximal twist. The details of the proof will be discussed in a forthcoming publication~\cite{CGJS}.

The short distance singularities on the r.h.s of Eq.~\eqref{specsum2} 
could correspond to additional $O(a)$ terms in the represenation
of the lattice correlator in the Symanzik effective theory.
More specifically these $O(a)$ are produced by the short distance expansion of two consecutive densities, since the short distance behavior of three or more densities leads to $O(a^2)$ or higher order terms. 

We can study the short distance singularities of a product of two operators through the operator product expansion (OPE). 
For a generic twist angle products like $P^+_{ab}(x)P^-_{bc}(0)$
will have an OPE for $x\rightarrow 0$ containing scalar densities.

Taking into account the presence of the contact terms and the standard Symanzik expansion 
we can write for the renormalized observable introduced in Eq.~(\ref{specsum2}) 

\begin{eqnarray}
\label{SymExp}
\sigma_{3,R}(\mu_R,m_R)
&=&-\int d^4x_1 d^4x_2 d^4x_3 d^4 x_4 d^4 x_5\expect{
   P^+_{12}(x_1)P^-_{23}(x_2)P^+_{34}(x_3)P^-_{45}(x_4)P^+_{56}(x_5)P^-_{61}(0) }_0 \nonumber\\
&&+a~{\rm S.T.} + a~{\rm C.T}\,.
\end{eqnarray}
where the densities on the r.h.s of Eq.~(\ref{SymExp})  are renormalized
operators that with abuse of notation we denote as the lattice
densities and $\expect{}_0$ represents the continuum
expectation values. The term labelled with S.T. corresponds to the standard terms
appearing in the Symanzik expansion and the one labelled with C.T.
corresponds to the $O(a)$ terms arising from the short distance singularities in the 
product of two densities. 
If we tune our lattice action parameters to achieve maximal
twist one can use the standard arguments leading to automatic $O(a)$ improvement to show that the S.T. vanish.

For the discussion of the C.T we keep generic values for the twisted and untwisted
quark masses. An example of the C.T. is given by

\begin{eqnarray}
\label{eq:ct}
&& \int d^4 x_2d^4x_3 d^4x_4 d^4x_5 \expect{S^{\uparrow}_{13}(x_2)P^+_{34}(x_3)P^-_{45}(x_4)P^+_{56}(x_5)P^-_{61}(0)}_0+\nonumber \\
&+&\int d^4 x_2d^4x_3 d^4x_4 d^4x_5 \expect{P^-_{23}(x_2)P^+_{34}(x_3)P^-_{45}(x_4)P^+_{56}(x_5)S^{\downarrow}_{62}(0)}_0\,,
\end{eqnarray}
where $S_{ac}^{\uparrow,\downarrow}=\psibar_a\frac{1}{2}(\id \pm\tau^3)\psi_c$\,.

We can now use non-singlet axial Ward-Takahashi identities to rewrite Eq.~\eqref{eq:ct} 
in a convenient form.
For twisted mass fermions at a generic twist angle we have

\begin{align}
\label{alltogetherWTI}
& \int d^4 x_2d^4x_3 d^4x_4 d^4x_5 \expect{S^{\uparrow}_{13}(x_2)P^+_{34}(x_3)P^-_{45}(x_4)P^+_{56}(x_5)P^-_{61}(0)}_0+\\
&+\int d^4 x_2d^4x_3 d^4x_4 d^4x_5 \expect{P^-_{23}(x_2)P^+_{34}(x_3)P^-_{45}(x_4)P^+_{56}(x_5)S^{\downarrow}_{62}(0)}_0 \nonumber\\
&=2m \int d^4 x_2d^4x_3 d^4x_4 d^4x_5 \int d^4x_1 \expect{P^+_{12}(x_1) P^-_{23}(x_2)P^+_{34}(x_3)P^-_{45}(x_4)P^+_{56}(x_5)P^-_{61}(0)}_0\,, \nonumber
\end{align}
where $m$ is the untwisted quark mass. All the other terms stemming
from the short distance singularities can be treated in an analogous
manner.  

\begin{figure}[t!]
\includegraphics[width=0.52\textwidth]{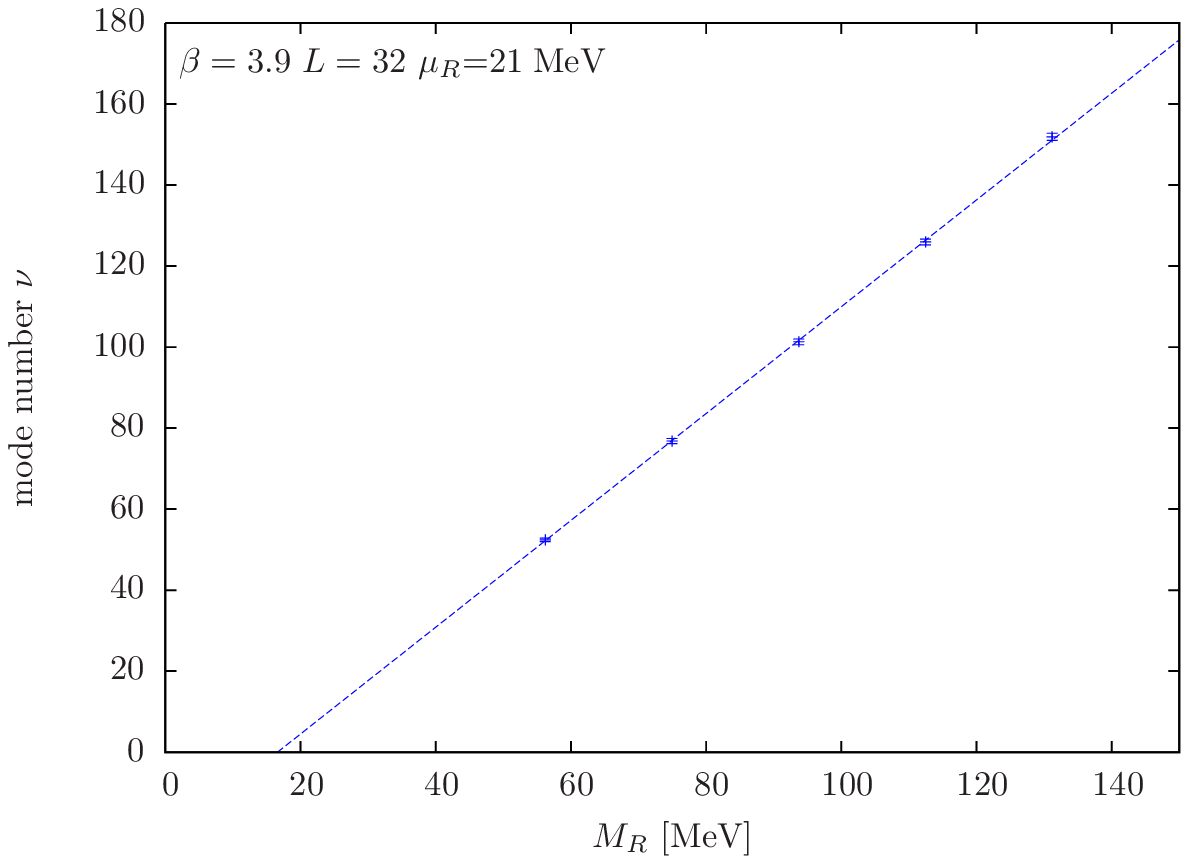}
\includegraphics[width=0.52\textwidth]{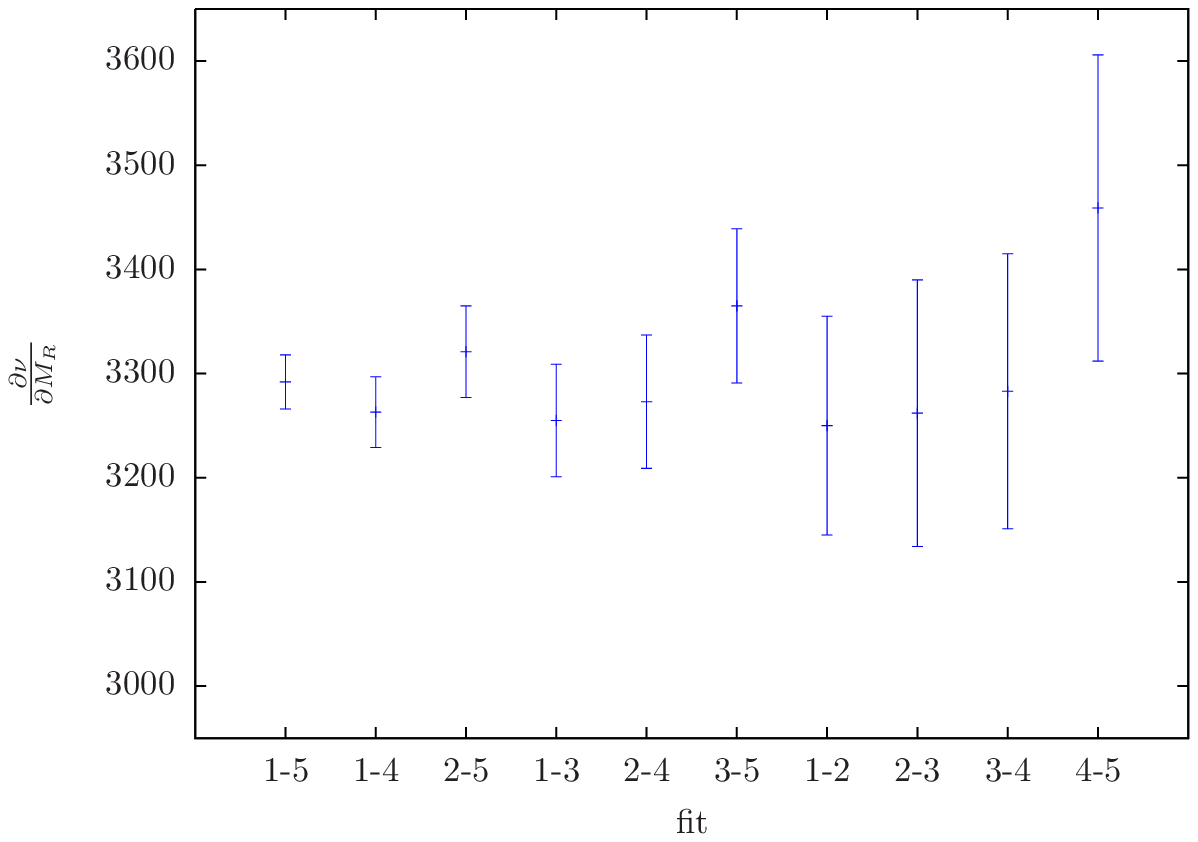}
\caption{(left) Mode number $\nu$ as a function of the renormalized
  threshold parameter $M_R$ for $N_f=2$ at $\beta=3.9$ and
  $a\mu=0.004$. The line corresponds to a linear fit to all 5
  points. (right) Result of the derivative $\partial \nu/\partial M$
  for different ranges in the linear fit. The x-axis represents the
  points included in the linear fit, where 1 corresponds to the lowest
  and 5 the largest value of $M$ respectively.}
  \label{fig:mn}
\end{figure}
For the sake of simplicity we have chosen to write a particular example for
six flavors, however, a generalization of this derivation for a
generic number of flavors is straightforward.

Our analysis indicates that despite the presence of additional $O(a)$ terms in 
density chain correlators, those terms turn out to be proportional to the untwisted
quark mass $m$ at a generic twist angle, thus they are bound to vanish at maximal twist.

\section{Chiral and continuum extrapolation of the chiral condensate}

In order to extract the chiral condensate through Eq.~(\ref{SigmaGL})
we computed the mode number using spectral projectors at several
values of the threshold parameter $M$ in a range from around 60 to 120
MeV. In this region a linear behavior is expected, which allows us to
compute the derivative which appears in Eq.~(\ref{SigmaGL}) from a
simple linear fit. Fig.~\ref{fig:mn}(left) shows the observed linear
behavior for a particular ensemble. 

\begin{figure}[t!]
\includegraphics[width=0.52\textwidth]{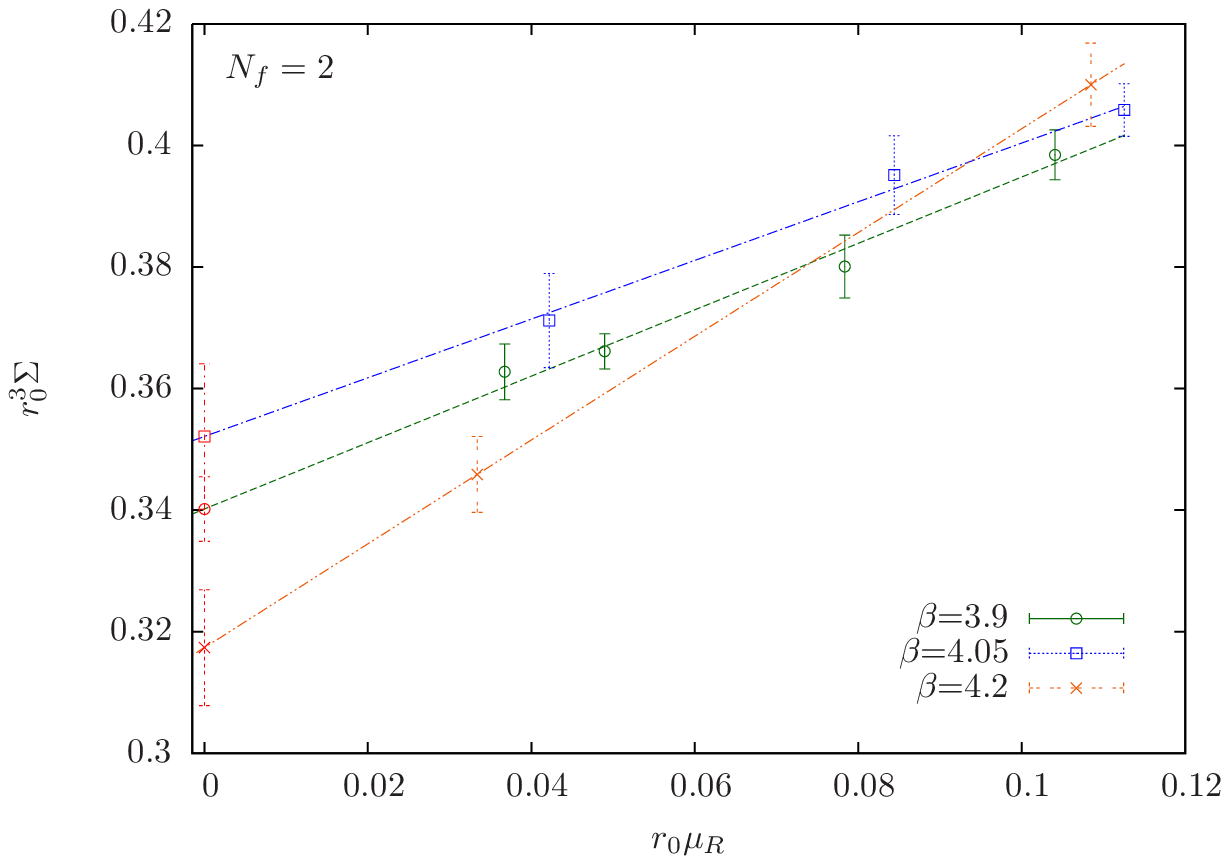}
\includegraphics[width=0.52\textwidth]{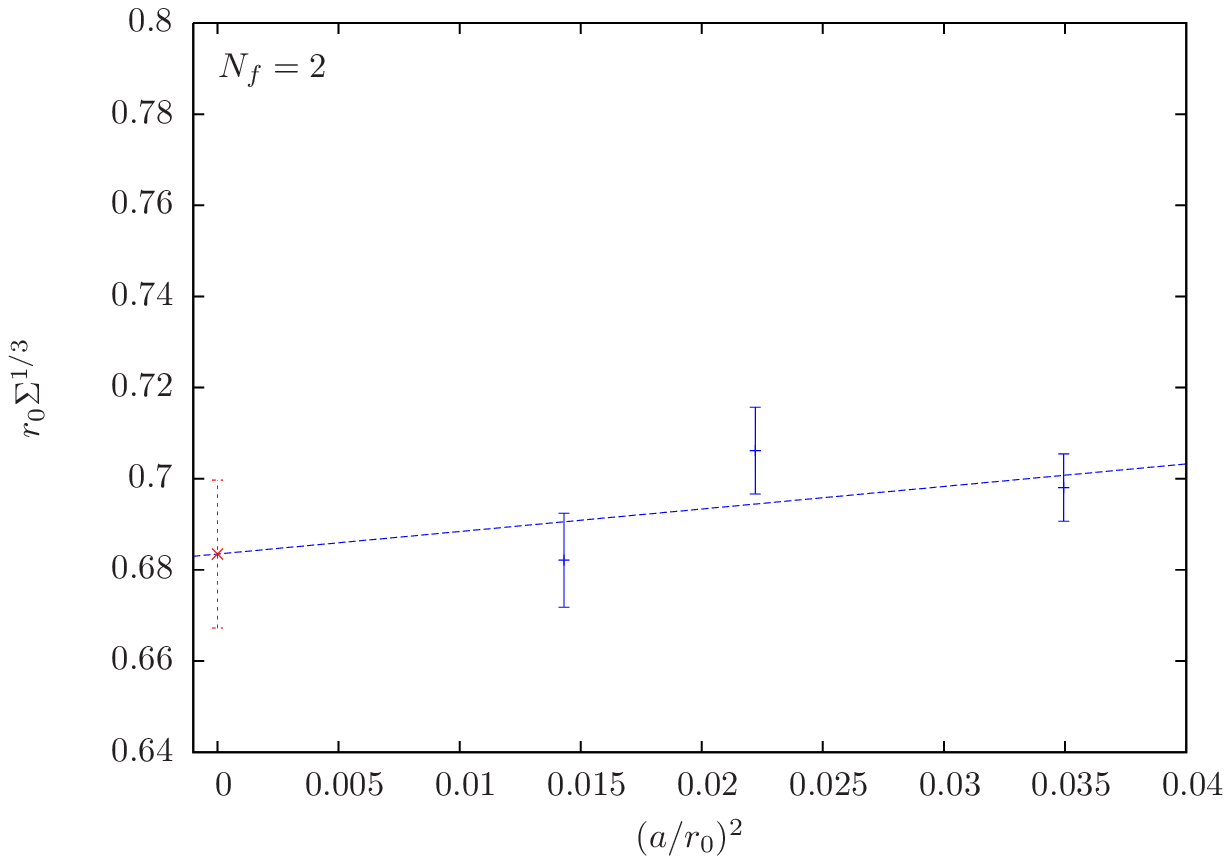}  
  \caption{Chiral extrapolation (left) of $r_0^3\Sigma$ for $N_f=2$ at three different values of the lattice spacing. Continuum limit (right) of chirally extrapolated $r_0\Sigma^{1/3}$ for $N_f=2$.}
  \label{fig:Nf2}
\end{figure}

To reliably estimate the correct range of $M$, where the mode number
behaves linearly, represents an important source of systematic
error in this calculation. We performed a comprehensive analysis of  the systematic
contributions to the error and concluded that the error coming from
the fitting range, is in fact the most prominent
\cite{Cichy:2013gja}. Recent studies \cite{Engel} found large
deviations from the linear behavior in their data and concluded that
NNLO effects could play an important role in the chiral extrapolation
of the condensate. In our case, however, with the available data, we
do not observe statistically significant deviations which shows that
such effects are very mild in our setup. 
To show this mild effect in Fig.~\ref{fig:mn}(right) the result of the derivative which
appears in Eq.~(\ref{SigmaGL}) for different ranges is shown, i.e. including
different points showed in Fig.~\ref{fig:mn}. One can see
an agreement in the result for the four lowest values of $M$ as
expected. The value of the mode number corresponding to the largest
$M$ seems to slightly deviate from the linear behavior. We consider
such deviation in our final systematic error. However, 
results from different fitting ranges are still statistically compatible. 

All the results presented in this section were computed using ensembles generated by the ETM collaboration. For further details we refer to the extended reference \cite{Cichy:2013gja} where all the relevant 
details are discussed.

\subsection{Results for $N_f=2$}

In this section we summarize the results of the chiral condensate for
$N_f=2$ dynamical fermions of maximally twisted mass fermions. The details of
the simulations can be found in Refs.~\cite{Baron:2010bv,Boucaud:2008xu}.

Fig.~\ref{fig:Nf2} (left) shows the chiral extrapolation of the
dimensionless ratio $r_0\Sigma$ for three different lattice spacings
which correspond to $a=0.085,\;0.067$ and $0.054$ fm respectively  \cite{Blossier:2010cr}. The
range of renormalized quark masses is from 15 to 45 MeV. We perform a
linear extrapolation to the chiral limit as suggested in 
Ref.~\cite{LuscherGiusti}, since the NLO effects are negligible for the 
mentioned ranges of quark masses and $M$, as we have explicitly tested.

The continuum limit of the chirally extrapolated condensate is shown
in Fig.~\ref{fig:Nf2} (right) and leads to the following result
$\quad r_0\Sigma^{1/3}=0.685 (16)(32)$, where the first error quoted combines the statistical
error and the error obtained from 
the uncertainty in the estimation of $Z_P$ and from $r_0/a$ in
quadrature. The systematic error corresponding to the uncertainty in
the linear regime of the mode number is given as the second error
quoted.

All the errors presented in this article were computed using the
method described in  Ref.~\cite{Wolff:2003sm}, whereas in 
Ref.~\cite{Cichy:2013gja} the bootstrap with blocking method was applied. Thus the
results slightly differ although they remain perfectly compatible. In
Ref.~\cite{Cichy:2013gja} the result is compared to others found in
the literature and a very good agreement is observed. 

\subsection{Results for $N_f=2+1+1$}

\begin{figure}[t!]
\includegraphics[width=0.52\textwidth]{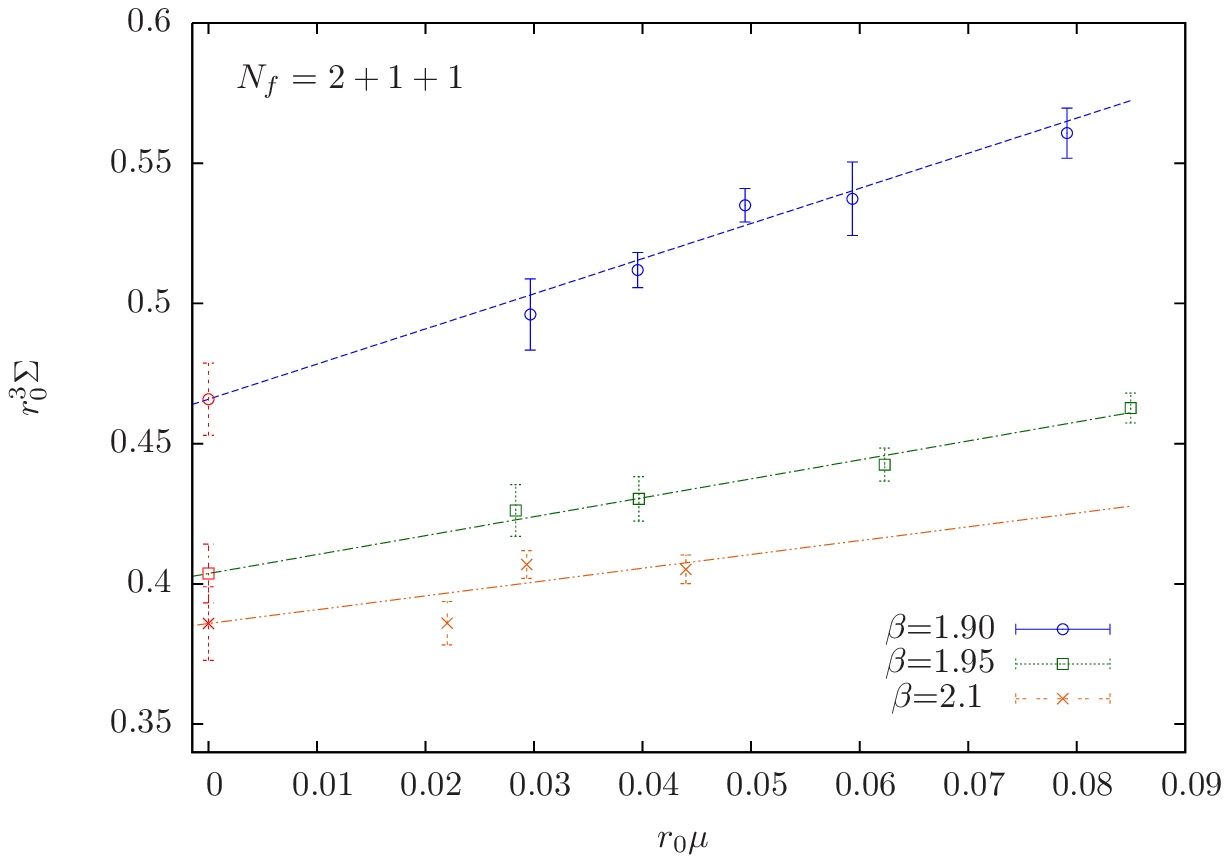}
\includegraphics[width=0.52\textwidth]{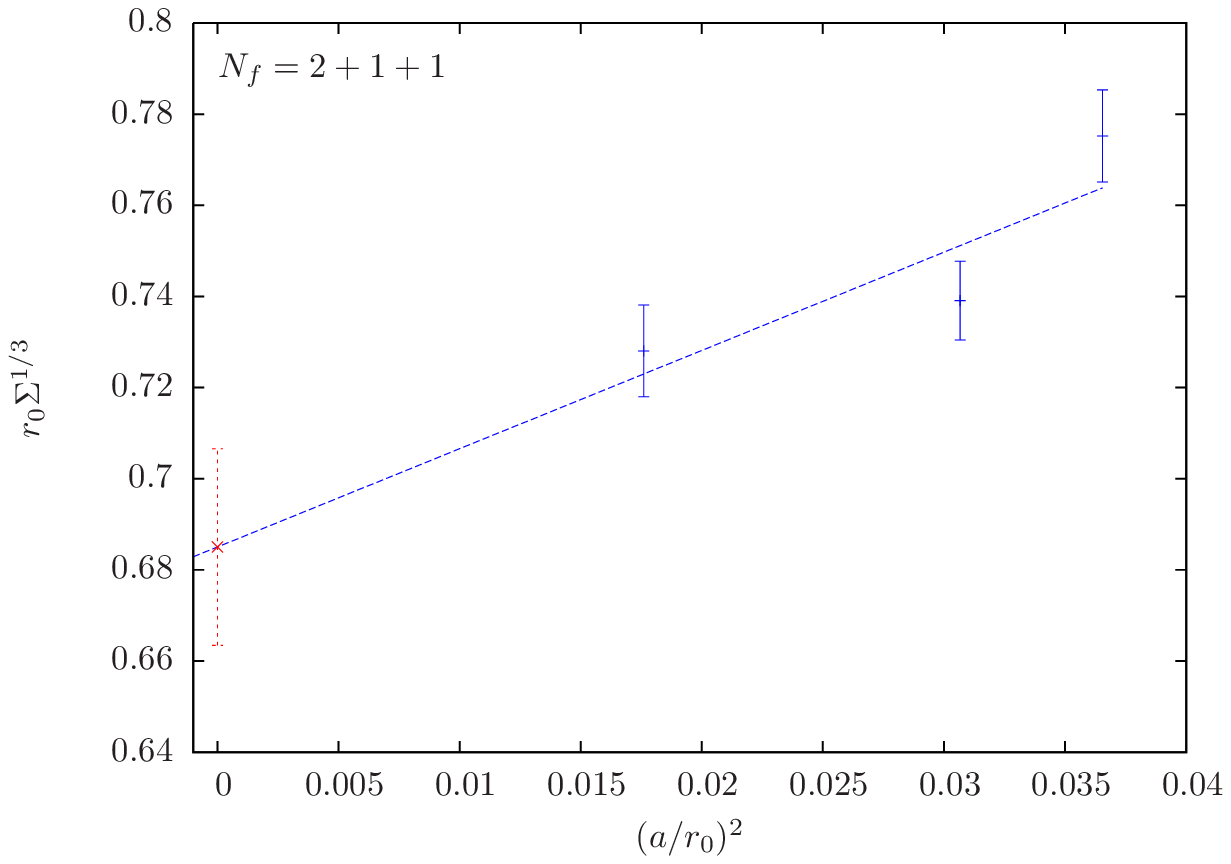}  
  \caption{Chiral extrapolation (left) of $r_0\Sigma$ for $N_f=2+1+1$ at three different values of the lattice spacing. Continuum limit (right) of chirally extrapolated $r_0\Sigma^{1/3}$ for $N_f=2+1+1$.}
  \label{fig:Nf211}
\end{figure}

In this section we summarize the results of the chiral condensate,
described in detail in Ref.~\cite{Cichy:2013gja}, for 
$N_f=2+1+1$ dynamical fermions of maximally twisted mass fermions
\cite{Baron:2010bv, Baron:2010th, Baron:2011sf}. 

Again we perform a chiral extrapolation following the strategy
presented in Ref.~\cite{LuscherGiusti} for three different values of
the lattice spacing $a=0.086, \;0.078$ and $0.061$ fm
respectively \cite{Baron:2011sf}. 
Fig.~\ref{fig:Nf211} (left) shows the results at
different values of the renormalized quark mass in a range between 13 and 45 MeV. The line represents the extrapolation at leading order of $\chi$PT.

In Fig.~\ref{fig:Nf211} (right) the continuum extrapolation is
plotted. Again we extrapolate in $a^2$ since, as we showed in the previous
section, the chiral condensate is $O(a)$ improved for twisted mass
fermions at maximal twist.

The final result for $N_f=2+1+1$ is $\quad$  $
r_0\Sigma^{1/3}=0.683(19)(18)$. The errors quoted represent the same uncertainties 
as in the result presented for $N_f=2$. Our value found for
$N_f=2+1+1$ flavours is compatible with the continuum  
limit value found for $N_f=2$ twisted mass fermions and quoted in the
previous section.

If we compare the errors of both, the result for $N_f=2$ and for
$N_f=2+1+1$, one can see that the data are less 
sensitive to the variations in the fit interval for $N_f=2+1+1$ than
for $N_f=2$.  This is mostly due to the fact that 
slopes for the case of $N_f=2+1+1$ flavors are smaller which 
contributes to decreasing the systematic error.

\section{Acknowledgements}
We thank the European Twisted Mass Collaboration for generating gauge
field configurations ensembles used for this work. 
E. Garcia-Ramos was supported by the Deutsche Forschungsgemeinschaft
in the SFB/TR 09. K.C. was supported by Foundation for Polish Science
fellowship ``Kolumb''. K. Jansen
was supported in part by the Cyprus Research Promotion Found
ation under contract $\Pi$PO$\Sigma$E$\Lambda$KY$\Sigma$H/EM$\Pi$EIPO$\Sigma$/0311/16. The numerical part of this project was carried out
in J\"ulich Supercomputing Center,  LRZ in Munich, the PC cluster
in Zeuthen and Poznan Supercomputing and Networking Center a(PCSS). We
thank these computer centers and 
their staff for all technical advice and help. We thank all
members of ETMC for useful discussions, in particular G. Herdoiza. We
also would like to thank L. Giusti and M. L\"uscher for insightful discussions.

\bibliographystyle{utphys}
\bibliography{bibliography}

\end{document}